\documentclass[english]{article}
\usepackage[T1]{fontenc}
\usepackage[latin9]{inputenc}
\usepackage{geometry}
\usepackage{verbatim}
\usepackage{float}
\usepackage{mathrsfs}
\usepackage{amsmath}
\usepackage{graphicx}  
\usepackage{authblk}
\usepackage{hyperref}

\makeatletter

\floatstyle{ruled}
\newfloat{algorithm}{tbp}{loa}
\providecommand{\algorithmname}{Algorithm}
\floatname{algorithm}{\protect\algorithmname}

\usepackage{babel}
\usepackage{pdfpages}

\makeatother

\usepackage{babel}
\begin{document}



# Simultaneous identification of specifically interacting paralogs and inter-protein contacts by Direct-Coupling Analysis

Thomas Gueudre[1,*], Carlo Baldassi[1,2,*], Marco Zamparo[1], Martin Weigt[3], Andrea Pagnani[1,2]

Affiliations:

[1] Department of Applied Science and Technology, Politecnico di Torino, 10129, Torino, Italy
[2] Human Genetics Foundation, Molecular Biotechnology Center, 10126 Torino, Italy
[3] Sorbonne Universités, UPMC, Institut de Biologie Paris-Seine, CNRS, Laboratoire de Biologie Computationnelle et Quantitative UMR 7238, 75006 Paris, France
Computationnelle et Quantitative UMR 7238, 75006 Paris, France
[*] T.G. and C.B. contribuited equally to this work.

Understanding protein-protein interactions is central to our understanding of almost all complex biological processes. Computational tools exploiting rapidly growing genomic databases to characterize protein-protein interactions are urgently needed. Such methods should connect multiple scales from evolutionary conserved interactions between families of homologous proteins, over the identification of specifically interacting proteins in the case of multiple paralogs inside a species, down to the prediction of residues being in physical contact across interaction interfaces. Statistical inference methods detecting residue-residue coevolution have recently triggered considerable progress in using sequence data for quaternary protein structure prediction; they require, however, large joint alignments of homologous protein pairs known to interact. The generation of such alignments is a complex computational task on its own; application of coevolutionary modeling has in turn been restricted to proteins without paralogs, or to bacterial systems with the corresponding coding genes being co-localized in operons. Here we show that the Direct-Coupling Analysis of residue coevolution can be extended to connect the different scales, and simultaneously to match interacting paralogs, to identify inter-protein residue-residue contacts and to discriminate interacting from noninteracting families in a multiprotein system. Our results extend the potential applications of coevolutionary analysis far beyond cases treatable so far.

# Introduction

Almost all biological processes depend on interacting proteins. Understanding protein-protein interactions is therefore key to our understanding of complex biological systems. In this context, at least two questions are of interest: First, the question *"who with whom"*, i.e. which proteins interact. This concerns both the networks connecting specific proteins inside one organism, however, in the context of this article, also the evolutionary perspective of protein-protein interactions, which are conserved across different species. Their coevolution is at the basis of many modern computational techniques for characterizing protein-protein interactions. The second question is the question *"how"* proteins interact with each other, in particular which residues are involved in the interaction interfaces, and which residues are in contact across the interfaces. Such knowledge may provide important mechanistic insight into questions like interaction specificity or competitive interaction with partially shared interfaces.

The experimental identification of protein-protein interactions is an arduous task, for reviews cf. [1,2]: High-throughput techniques that aim to identify protein-protein interactions *in vivo* or *in vitro* are well documented and include large-scale yeast two-hybrid assays and protein affinity mass spectrometry assays. Such large-scale efforts have revealed useful information, but are hampered by high false positive and false negative error rates. Structural approaches based on protein co-crystallization are intrinsically low-throughput and of uncertain outcome due to the unphysiological treatment needed for protein purification, enrichment and crystallization.

It is therefore tempting to use the exponentially increasing genomic databases to design in-silico techniques for identifying protein-protein interactions, cf. [3,4]. Prominent techniques to date include the search for co-localization of genes on the genome (e.g. operons in bacteria) [5,6], the Rosetta-stone method (domains fused to a single protein in some genome, are expected to interact in other genomes) [7,8], but also co-evolutionary techniques like phylogenetic profiling (correlated presence or absence of interacting proteins in genomes) [9] or similarities between phylogenetic trees of groups of orthologous proteins (cf. the mirrortree method) [10,11]. Despite success of all these methods, their sensitivity is limited due to the use of relatively coarse global criteria (genomic location, phylogenetic distance) instead of full amino-acid sequences.

The availability of thousands of sequenced genomes [12] thanks to next-generation sequencing techniques enables the application of much finer-scale statistical modeling approaches, which take into account the full sequence [13]. In this context, the *Direct-Coupling Analysis* (DCA) [14] had been developed to detect direct inter-protein coevolution and, in turn, inter-protein residue-residue contacts between bacterial signal-transduction proteins, to help to assemble protein complexes [15,16] and shed light on interaction specificity [17,18]. The applicability of DCA and related coevolutionary approaches [19,20] to protein-protein interactions far beyond signaling system has been recently established [21,22].

However, these methods require a large joint MSA of at least about 1000 amino-acid sequence pairs to work accurately. Each line of this MSA concatenates a *pair of interacting proteins*. So far

the application of coevolutionary methods remains therefore restricted to those cases, where such joint alignments could be constructed easily:
- Each species has only a single copy of the family, i.e. no paralogs exist. Matching of interacting proteins can be achieved by *uniqueness in the genome.*
- Even if paralogs exist, genes of interacting proteins are frequently co-localised on the genome and can therefore be matched by chromosomal vicinity. This is true in particular in the case of bacteria; functionally related proteins are frequently coded for in operons and consequently co-transcribed.

Co-localisation is used extensively in the construction of joint MSA for covariance analysis [22,23,24,25]. However, the case of multiple paralogs with non co-localised genes has remained out of reach for coevolutionary analysis, despite its enormous relevance: Out of the 4499 Pfam-29 [26] protein families with more than 500 sequences, 3221 have on average more than 2 paralogs per species, and 1378 families even more than 5 paralogs. Another observation underlines the importance of addressing generally localized genes: Out of 3643 protein-protein interactions reported for *Escherichia coli* in the IntAct Molecular Interaction database [27], only 1341 (36.8%) concern intra-operon interactions.

Here we suggest a novel approach to solve the problem of matching paralogs, which is based on a *simultaneous construction of the joint MSA and detection of inter-protein coevolution.* In a nutshell, the method is based on the idea that the correct matching of interacting paralogs maximises the inter-protein coevolutionary signal. The corresponding optimization problem turns out to be extraordinarily hard to solve exactly. We therefore propose two approximate strategies: The first one is computationally very efficient and of sufficient accuracy for subsequent contact prediction. If interaction partner prediction is the central task, a slower but more accurate iterative scheme can be used. The validity of the approach is demonstrated in the cases of bacterial two-component signal transduction and the protein-protein interaction network between the proteins of the Tryptophan biosynthesis pathway (Trp pathway). Our findings open the field to broad applications to protein interactions beyond single-copy or co-localised protein-coding genes, and help to bridge the multiple scales of inter-protein coevolution.

## Results

### An efficient approach to paralog matching in interacting protein families

*Paralog matching by maximising the inter-family covariation*

In this paper, we show that DCA may help to solve the before-mentioned paralog problem by simultaneously matching paralogs and determining inter-protein coevolutionary scores. We argue that the best matching is actually the one maximising the inter-protein covariation, empirical evidence for the correctness of this idea will be provided later in this Results section.

Let us formalize the task (cf. Appendix and the S1 Supp. Inf. for details): Given are multiple-sequence alignments for two protein families, denoted as $X^1 = \{a_i^m \,|\, i = 1, \dots, L_1, m = 1, \dots, M_1\}$

and $X^2 = \{b_j^n \mid j = 1, ..., L_2, n = 1, ..., M_2\}$. Each row is the aligned sequence of a protein of length $L_1$ (resp. $L_2$), each column a specific homologous position in the protein. The proteins belong to $S$ species having $M_1^1, ..., M_1^S$ paralogs in protein family 1 (analogous notations are used for protein family 2), with $M_1 = M_1^1 + \cdots + M_1^S$ being the total alignment depth. Proteins belonging to species $k$ have indices $I_1^k$. A paralog matching of this species is thus a (partial) mapping $\pi_k: I_1^k \to I_2^k$ of proteins from the first to the second family. The total paralog matching is given by the mapping $\pi = (\pi_1, ..., \pi_S)$ of all species, it generates a joint alignment $(X^1 \circ X^2)_\pi$ by concatenating all matched protein pairs into rows of length $L = L_1 + L_2$. We consider an *injective matching strategy* described in Panel A of Fig. 1: For each species, all proteins from the family with lower paralog number are matched to pairwise different proteins in the other family. In this article we do consider neither the sparse case, where only part of the sequences are matched, nor cases of promiscuous interaction, where one protein should be matched to several others.

For a given matching $\pi$, i.e. for a given joint MSA $(X^1 \circ X^2)_\pi$, we can calculate the amino-acid covariance matrix $C^\pi$, which has dimension $20L \times 20L$, cf. Appendix. Within the Gaussian modeling approach introduced in [28], the normalized log-likelihood of the model is given by $L^\pi = -\frac{1}{2} \log \det C^\pi$. The best matching will maximise the log-likelihood, $\pi^* = argmax_\pi(L^\pi)$. However, due to the huge number of possible matchings, which is exponential in the number of species, and super-exponential in the number of paralogs inside each species, the exact solution of this optimization task is infeasible. Furthermore, we empirically observed this discrete optimization problem to be plagued by many local likelihood maxima, such that local algorithms easily get stuck.

We propose two heuristic algorithms to approximate the solution of this optimization task. A fast progressive method is applicable to large-scale data sets (e.g. many pairs of large families). While having limited accuracy in identifying specifically interacting paralogs, it is suitable for subsequent inter-family DCA analysis to predict residue-residue contacts between proteins, or to discriminate interacting from non-interacting families. A slow but accurate iterative method is more suitable for smaller-scale problems, where the accurate identification of individual interacting protein pairs is central.

### *An efficient progressive paralog matching (PPM) algorithm*

A first algorithmic strategy to find the matching maximizing the inter-family covariation is inspired by progressive techniques in constructing multiple-sequence alignments [29]: Species are matched progressively, starting with the simplest ones (species with low paralog numbers in our case) and iteratively adding more complicated species with higher paralog numbers. Each species is matched only once, on the basis of all already matched species. Our *progressive paralog matching* (PPM) algorithm proceeds as follows, technical details are provided in Appendix and in the S1 Supp. Inf., the pipeline is depicted in Fig. 1B:

1. Species are ordered according to the entropy of their possible matchings, i.e. to the expected hardness of the matching task.
2. Species of low entropy are used to generate a seed matching, in our specific case zero-entropy species, i.e. species with a single paralog are used.

3. In order of increasing entropy, species are added recursively: *(a)* DCA is applied to the already matched MSA. *(b)* The DCA parameters are used to score all pairs of paralogs inside the new species to be added. *(c)* An optimal matching for the new species is constructed using these scores.

The algorithm terminates when all species are included. The absence of iterative error correction makes this algorithm computationally very fast. However, early on fixed errors may propagate through the whole procedure and disturb later on matched species.

*An accurate iterative paralog matching (IPM) algorithm*

The PPM algorithm does the matching of the proteins belonging to each species only once, based only on previously matched species. Any matching error made at some stage is kept up to the end, causing possibly other matching errors. It would be possible to correct at least part of these errors when considering later included proteins. However, the likelihood landscape is very rough and found to have many local maxima, so a simple iterative refinement remains stuck close to the PPM result.

To overcome this limitation, our slow but accurate *iterative paralog matching* (IPM) algorithm follows three steps:
1. Generate K random paralog matchings respecting species. In practical applications, K=256 was found a good compromise between computational time and accuracy.
2. Independently refine all K matchings iteratively by hill climbing (discrete analogous of gradient ascent), see Appendix and the S1 Supp. Inf. for a detailed description, until convergence to a local likelihood maximum.
3. Merge pairs of matchings by using average DCA scores; subsequently refine the merged consensus matching by hill climbing. Iterate until a single matching is left. A final refinement step is described in S1 Supp. Inf.

The idea behind the merging step is simple: The consensus of two imperfect matchings should reinforce the common signal as compared to the random noise. This non-local change of the matching is found to be able to escape local log-likelihood maxima, which otherwise trap local algorithms like hill climbing. Details of the algorithm are given in Appendix and in the S1 Supp. Inf., the pipeline is depicted in Fig. 1C.

## Simultaneous identification of interaction partners and inter-protein residue-residue contacts in bacterial signal transduction

To test both algorithmic strategies, we first consider bacterial two-component systems (TCS) [14], which are the most diffused signal-transduction systems in the bacteria. TCS have played a prominent role in the development of DCA [14]. They consist of two interacting proteins, the Histidine sensor kinase (SK) as a signal receiver, and the response regulator (RR), which under activation typically acts as a transcription factor and triggers a transcriptional response. In particular we use the dataset of Procaccini et al. [17], which collects 8998 interacting protein pairs from 712 distinct species, cf. Appendix. A random matching between SK and RR inside species would make, on average, one correct prediction per species, i.e. only a fraction of 712/8998 = 7.9% of all matched SK/RR pairs would be correct. Earlier approaches to match SK

and RR have used Bayesian residue networks [23] or aligned protein-similarity networks [31]: While they improve substantially over random matchings, their accuracy remains inferior to the algorithms presented below.

We first check the self-consistency of our matching idea: Is our MSA of SK/RR, which are co-localized in joint operons and therefore expected to be truly interacting, stable under the matching procedure? To answer this question, we infer a DCA model using this MSA, and rematch all species. No changes are observed: The true MSA is actually a fixed point of the proposed algorithmic procedure.

As second step, we run PPM. Only 59 out of 8998 sequence pairs are matched immediately because both SK and RR are unique in the genome. The extension of this seed matching by PPM is shown in Fig. 2A: While the seed matching is insufficient to predict inter-protein contacts between SK and RR (only one true contact out of the best 15 inter-protein predictions), it is sufficient to guide PPM to 84.7% precision (7620 out of 8998 cognate pairs are correctly identified). The final matching is sufficient to provide accurate inter-protein contact predictions, all of the 15 highest-scoring residue pairs are true inter-protein contacts (distance 8Å in PDB 3dge [32]). We observe that, with increasing size of the progressive matching, the contact prediction becomes more and more accurate: For 1014 matched sequences, 10 out of the first 15 DCA predictions are inter-protein contacts, for 2000 matched sequences even 13 out of 15, cf. Fig. 2B for a more quantitative assessment. Beyond contact prediction, Fig. 3A shows that the highest inter-protein scores in the list of matched proteins exclusively indicate truly interacting pairs. All of the first 1347 pairs are interacting.

While PPM is computationally very efficient, its accuracy in identifying true interaction partners is limited. This results from the progressive strategy: Once a matching error is made, it is not corrected but influences all subsequently matched species. To this end we have applied the computationally more involved IPM algorithm, cf. Fig. 3C: While the hill climbing steps reach an accuracy comparable to PPM, the non-local merging steps reach 91.2% of precision (8206 true matches). The IPM allows also for testing our ground hypothesis that there is a correspondence between the log-likelihood of a matching and its accuracy. While IPM proceeds to maximize the log-likelihood, the matching error is, up to fluctuations, monotonously decreasing. Furthermore, the insert of the Fig. 3C shows that IPV slightly exceeds the log-likelihood of the true operon-based matching, but the error rate is not decreasing any more beyond that point. This suggests that the intrinsic error rate of the association between log-likelihood and matching error is close to 9%.

## Simultaneous identification of interacting families and specifically interacting proteins in a bacterial metabolic pathway

DCA has been used to identify interacting protein families [21,25]. Based again on the availability of large joint MSA, only pairs of families showing significant inter-protein coevolution are expected to interact. Again, we argue that even without a large known set of (potentially)

interacting protein pairs, the PPM strategy simultaneously creates such an alignment, and the inter-protein score is informative about inter-family interaction.

To demonstrate this, we use the Tryptophan biosynthesis pathway comprising seven different proteins TrpA-TrpG, which catalyze subsequent reactions in the pathway. Among the 21 protein pairs, only two are known to interact based on experimentally resolved co-crystal structures: TrpA-TrpB (PDB 1k7f [33]) and TrpE-TrpG (PDB 1qdl [34]). While individual Pfam MSA sizes reach from 8,713 sequences for TrpF to 78,265 for TrpG, pairing by uniqueness in the genome only in three cases leads to joint alignments beyond 1000 sequences (TrpC-TrpF 1578, TrpA-TrpC 1546, TrpA-TrpF 1433), while the actually interacting pairs have extremely small joint MSA of 8 sequences for TrpE-TrpG and 95 sequences for TrpA-TrpB. No detection of interactions is possible with such small alignments, cf. Fig. 4. In [35] we have shown that matching by genomic colocalization leads to joint alignment size of 2519-8053 sequences, with a majority below 4000 sequences. These alignments separate the two known interacting pairs (DCA scores 0.3, 0.38) from an almost continuous background of scores up to 0.17.

To test our paralog matching, we apply PPM to each of the 21 Trp protein pairs, cf. Fig. 4. The seed matchings, generated by uniqueness in the genome, range from 8 to 1578 protein pairs. They do not allow for recovering the correct interacting family pairs (ranks 5 and 21 out of 21). When having matched 1000 protein pairs in each family, the three highest-scoring protein pairs are TrpA-TrpB (score 0.23), TrpF-TrpG (score 0.18) and TrpE-TrpG (score 0.17), followed by almost continuous scores up to 0.15. The correct interactions thus have ranks 1 and 3, but no gap to scores of non-interacting pairs exists.

Using the full progressive matchings, TrpA-TrpB has score 0.34 and TrpE-TrpG 0.25, followed by almost continuous scores up to 0.15. The two correct interactions are recognized with a gap, which is almost as large as in the matching obtained using genomic colocalization, illustrating again the strong capacity of our method to recover accurately the matching between interacting proteins.

Results obtained at the level of interaction networks can be corroborated by inter-protein contact predictions obtained for the two interacting pairs, cf. Fig. 4E-F: For TrpA-TrpB, 9 out of the first 10 interprotein-contact predictions are true positive, and 12 out of the first 15. The situation is very similar for TrpE-TrpG: 10 out of the first 10, 11 out of the first 15 predicted pairs are in contact across the interface.

This shows that, in the case of the Trp pathway proteins, the progressive paralog matching strategy is able to create large enough joint alignments for pairs of families, which allow to (i) distinguish interacting from non-interacting families, and (ii) to predict inter-protein contacts for the interacting ones.

## Discussion

Global methods to detect coevolution, like DCA, PsiCov and GREMLIN, have recently enjoyed growing popularity in a very specific setting: Starting from a large multiple sequence alignment of homologous proteins, these approaches have helped to extract residue-residue contacts from residue-residue amino-acid covariation. In the context of interacting proteins, the inferred inter-protein contacts have in turn helped to structurally assemble protein complexes.

However, the applicability of these methods has remained limited due to the *a priori* need to obtain joint multiple sequence alignments of pairs of interacting proteins, with each row containing a pair of interacting proteins out of two protein families. This MSA has to be obtained by *external information* like the uniqueness of the two protein families inside a species (no paralogs present) or the genomic co-localization in bacterial operons.

In this work, we show that one can turn the argument around: *The coevolution between two protein families itself can be used to identify interacting partner proteins*, and thereby to generate the joint MSA simultaneously with obtaining an inter-protein contact prediction. We have shown that an accurate matching between proteins families can be obtained, which (i) connects only proteins in the same species and (ii) maximizes the detectable inter-family coevolutionary signal. The idea is that basically any mismatch connecting two non-interacting proteins, decreases the inter-family covariation. In Fig. 3 we have actually observed that there is, up to a statistical fluctuations, a monotonously decreasing relation between the log-likelihood of a matching (which is a measure of the total inter-family coevolutionary signal) and the error rate in the matching, when compared to a Gold-standard data set of co-localized bacterial proteins from two-component signal transduction pathways. However, once the log-likelihood of the Gold-standard matching was obtained (or even slightly exceeded), the residual matching error of about 9% did not decrease any more. This may be a sign for an intrinsic limitation of the idea connecting likelihood and matching accuracy, but it may also be biological signal. About 60% of the mismatched were pairwise switches (transpositions) between two TCS, 18% concern triples. It has been speculated before, that 15-20% of all bacterial signaling systems display some tendency to crosstalk, i.e. interactions are not really one-to-one. Part of the "mismatched" proteins could actually been read as predictions for inter-TCS crosstalk, however, in model species *Escherichia coli* and *Bacillus subtilis*, where cases of crosstalk have been reported [36,37], no matching errors were found. Experimental tests in other species would be needed to test this hypothesis.

The intuitive idea of maximizing the inter-family coevolutionary signal leads, however, to a computationally extremely hard problem: The search space (i.e. all possible joint MSA) is exponentially large in the number of species and super-exponential in the number of paralogs inside each species. The problem would become much simple to solve if the log-likelihood – based on a *global* modeling of the sequence variability – could be replaced by a *local* correlation measure maximizing, e.g., the Frobenius norm of the inter-protein covariance matrix. This is implemented as a first fast stage of the iterative paralog matching algorithm, but in the case of TCS it gets stuck at a high error rate of almost 40% of mismatches. *Global modeling is*

*necessary to reach high accuracy in paralog matching*. We have also seen that the accuracy drops only slightly (error rate ~15%) when the slow iterative procedure is replaced by a fast progressive paralog matching. The resulting joint alignments are sufficiently precise to enable accurate inter-protein contact prediction, and, as illustrated in a bacterial metabolic pathway, to discriminate between interacting and noninteracting protein families.

The two strategies – progressive and iterative paralog matching – open both the road to large-scale analysis for predicting currently unknown protein-protein interactions based on inter-protein coevolution, and more detailed smaller-scale studies in the structural biology of interacting proteins. As stated previously, coevolution based procedures to analyze PPI have extensively used co-localization [22,23,24,25]. A natural question is what fraction of the known bacterial interactome comes from co-localized genes? Given our partial knowledge of the interactome at present, we still cannot provide a precise answer to this question. However, we can give a partial estimate based on current knowledge in *E. coli*, i.e. in the currently best-studied model species. Its proteome consists of 4323 non-redundant proteins organized in 2148 operons (817 of which host at least two genes). This results in 4885 potential PPIs within the same operon, in comparison to more than 9 million protein pairs in total. IntAct [27], one of the most comprehensive database for PPI network, reports 3643 PPI for *E. coli*, of which more than one third (1341 pairs) are intra-operon PPI. iPfam [38], a domain-based database of structurally known PPI, reports 4100 interacting family pairs (~2000 of which are homodimers). The breakdown of the 4885 possible intra-operon interactions in terms of distinct protein domains, gives 8068 distinct intra-operon domain pairs. Of the 2100 heterodimeric domain pairs in iPfam, only 640 are present in *E. coli*, 214 of which are in the same operon. Again, despite the difference of the experimental resources reported in the two databases, about one third of the known interactions, originate from the same operon. Our methodology provides an efficient and scalable algorithmic strategy to analyze the remaining two third of the known interactome, for which criteria such as genomic proximity cannot be used. For this reason we believe that our development of an information theory based approach to match members of homologous protein families by maximizing the inter-alignment information is a fundamental step to unravel protein interactome at large.

## Author contributions

T.G., C.B., M.W., A.P. designed research, T.G., C.B., M.Z., M.W., A.P performed research, T.G., C.B. analyzed data, M.W., A.P. wrote the paper.

## Note

While finalizing this manuscript, we learned that AF Bitbol, RS Dwyer, LJ Colwell and NS Wingreen are preparing a related paper on predicting interacting paralog pairs.

# Appendix

## Gaussian Direct Coupling Analysis

The basis of the paralog matching procedure is the Gaussian Direct Coupling Analysis formulated in [28]. Let us assume a matched MSA $A$ of $M$ sequences of length $L$. The MSA is transformed into a $M \times 20L$-dimensional binary array $X$ by replacing each amino acid by a distinct 20-dimensional vector containing one entry "1" and nineteen entries "0"; gaps are represented by zero-vectors. The empirical covariance matrix of the transformed MSA is the $20L$-dimensional square matrix $C$ (the explicit dependence on the matching leading to the MSA is suppressed here), the empirical mean is the $20L$-dimensional vector $\mu$, cf. the S1 Supp. Inf. for the precise definition of these quantities using standard DCA sequence weighting and pseudocounts. Given these empirical matrices the Gaussian DCA model assigns a probability

$$P_G(x \mid \mu, C) = \frac{1}{\sqrt{(2\pi)^N det(C)}} exp\left[-\frac{1}{2}(x-\mu)^T C^{-1}(x-\mu)\right],$$

to any amino-acid sequence of length $L$ in binary representation. From this expression, the log-likelihood of the original MSA $X$ can be easily determined as $\mathcal{L} = log[P_G(X \mid \mu, C)] = -1/2 \, log \, det \, C$, cf. the S1 Supp. Inf. Our matching strategy aims at maximizing this likelihood by selecting the matching leading to the joint MSA $X$. For computational reasons, we will also use, as a faster to compute scoring function, the squared Frobenius norm of the covariance matrix $C$, i.e. $||C||_F^2 = \sum_{i,j} C_{i,j}^2$.

The two matching strategies are described in the Results section, extensive details are provided in the S1 Supp. Both PPM and IPM require an optimal assignment within each species from the complete set of pairwise scores (either log-likelihood, or Frobenius norm) computed from the inferred model. As the optimal assignment problem can be easily formulated in terms of linear programming, we used the Gurobi library [39] to efficiently solve it.

## Data extraction

**Two Component systems**: The data for the SK/RR analysis were originally published in [17], here we give a short description: 769 bacterial genomes were scanned using hmmer [40] with the Pfam 22.0 Hidden Markov Models [43] for the following Sensor Kinase (SK) domains: "HisKA" (PF00512), "HWE_HK" (PF07536), "HisKA_2" (PF07568), "HisKA_3" (PF07730), "His_ kinase" (PF06580), and "Hpt" (PF01627), and for the Response Regulator (RR) domain "Response_reg" (PF00072). Using a simple operational definition of an operon as a sequence of consecutive genes of same coding sense, and with intergenic distances not exceeding 200 base pairs, a total M = 8,998 SK/RR pairs were identified in operons containing a single SK (of type HisKA) and a single RR domain. As reference structure we consider the PDB entry 3dge [32].

**Trp operon:** The tryptophan biosynthetic pathways consists of 7 enzymes (TrpA,B,C,D,E,F,G). Only two protein-protein interactions are known and resolved structurally: TrpA-TrpB (PDB 1k7f

[42]) and TrpG-TrpE (PDB 1qdl [43]). Single-protein MSA have been extracted using the pipeline proposed in [25]: (i) Extract sequences corresponding to names from Uniprot; (ii) Run MAFFT [44] using: `mafft --anysymbol --auto`; (iii) Create a profile Hidden Markov Model using hmmbuild from the hmmer suite, and search Uniprot using `hmmsearch` [45]; (iv) Remove inserts.

FIGURES

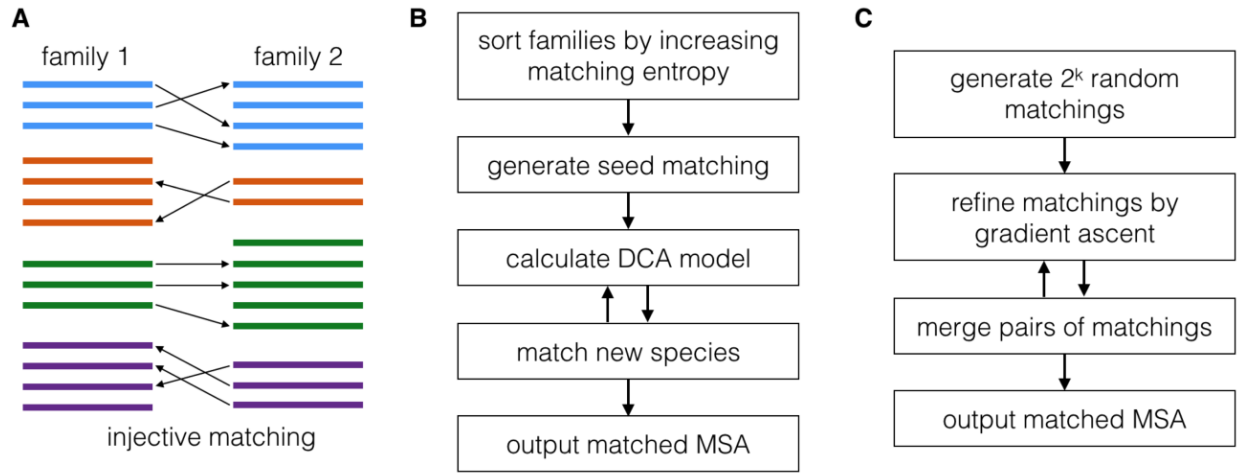

**Figure 1: Paralog matching procedures. Panel A** shows the considered injective strategy to match paralogs. For each species (depicted by different colours), each paralog from the species with the lower paralog number is matched to a distinct sequence in the other species. **Panel B** shows the pipeline of the progressive paralog matching (PPM) algorithm. Species are sorted by increasing matching entropy. Starting from a seed matching (generated in our case by uniqueness in genome), the algorithms calculates the DCA model, uses it to add and match a new species, and iterates these two steps until all species are matched. **Panel C** shows the iterative paralog matching (IPM) pipeline. $2^k$ random matchings are generated, each one is independently refined using hill climbing of the likelihood. After refinement, pairs of matchings are merged using average matching scores. Refinement and merging are iterated until only a single refined matching is left.

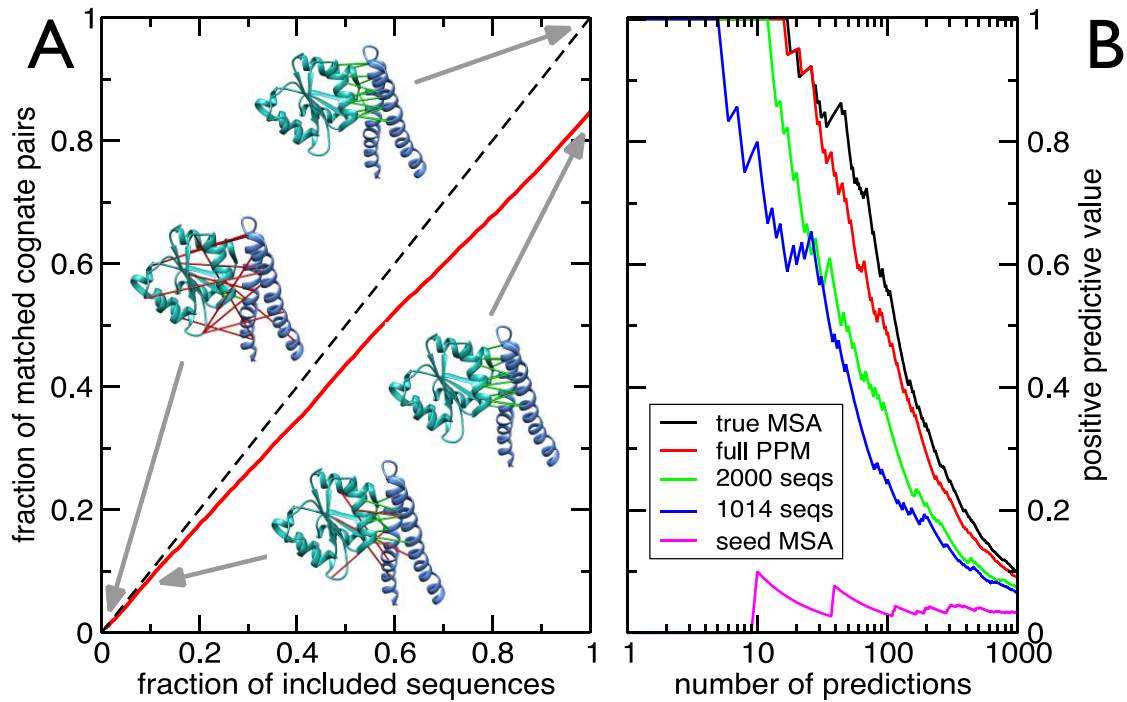

**Figure 2: The progressive matching procedure matches cognate pairs and enables inter-protein residue-contact prediction. Panel A:** The red line shows the fraction of the 8998 cognate pairs, which are correctly matched by the progressive matching algorithm, as a function of the matched pairs. A perfect matching procedure would follow the dashed diagonal. The SK/RR complex structure is overlaid with the 15 highest-scoring contact predictions, for the seed alignment, after having matched 1014 proteins, and at the end of the matching. Green bonds show correct, red incorrect predictions (contact cutoff 8Å). The upper structure shows the prediction obtained with the full cognate MSA.
**Panel B:** The positive predictive value (i.e. the fraction of true positives in between all inter-protein contact predictions) is shown as a function of the number of predictions, for several joint MSA (the true operon-based cognate matching (black), and successive states of the PPM for the seed alignment (magenta) and 1014 (blue), 2000 (green) resp. 8998 (red). The prediction accuracy grows during the progressive matching, and finally reaches almost the accuracy of the cognate matching.

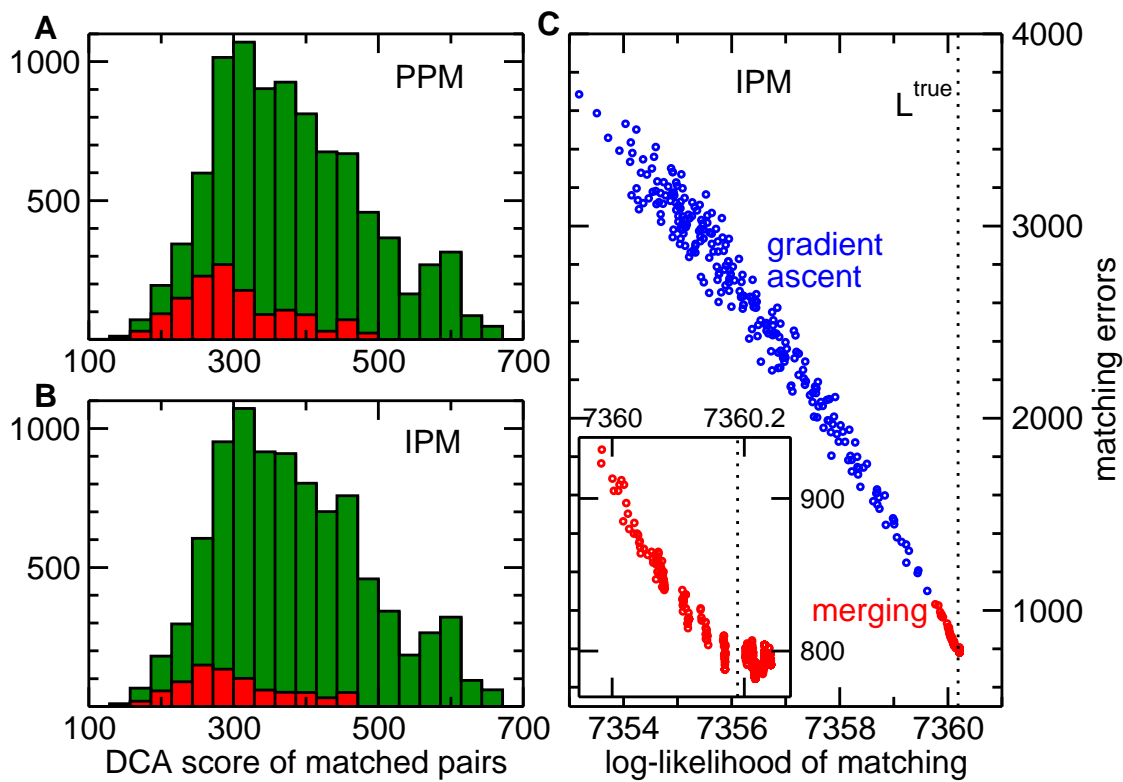

**Figure 3: Detailed performance of the PPM and IPM algorithms. Panel A (resp. Panel B)** shows the histogram of DCA scores of the final PPM (resp. IPM) matching. The fraction of true positive (TP) predictions is colored in green, the fraction of false positive (FP) predictions in red. While the high-scoring pairs are exclusively TP, low and intermediate scores show a mixture of TP and FP. The overall histogram is only insignificantly shifted towards higher scores when comparing IPM to PPM, but the overall weight of the FP is visibly decreased. **Panel B** shows the dependence of the number of matching errors (FP) on the log-likelihood of the IPM matching. Iteration proceeds from the upper left to the lower right corner, first by hill climbing of the log-likelihood (blue points), then by merging locally optimal matchings (red point). The overall procedure arrives at a log-likelihood which is slightly superior to the one of the true matching (dotted vertical line), at a precision of about 91.2% (8206 TP out of 8998 TP+FP). The insert enlarges the final steps of the merging procedure, it becomes evident that the almost linear relation between log-likelihood and error breaks down once the log-likelihood of the cognate matching is reached.

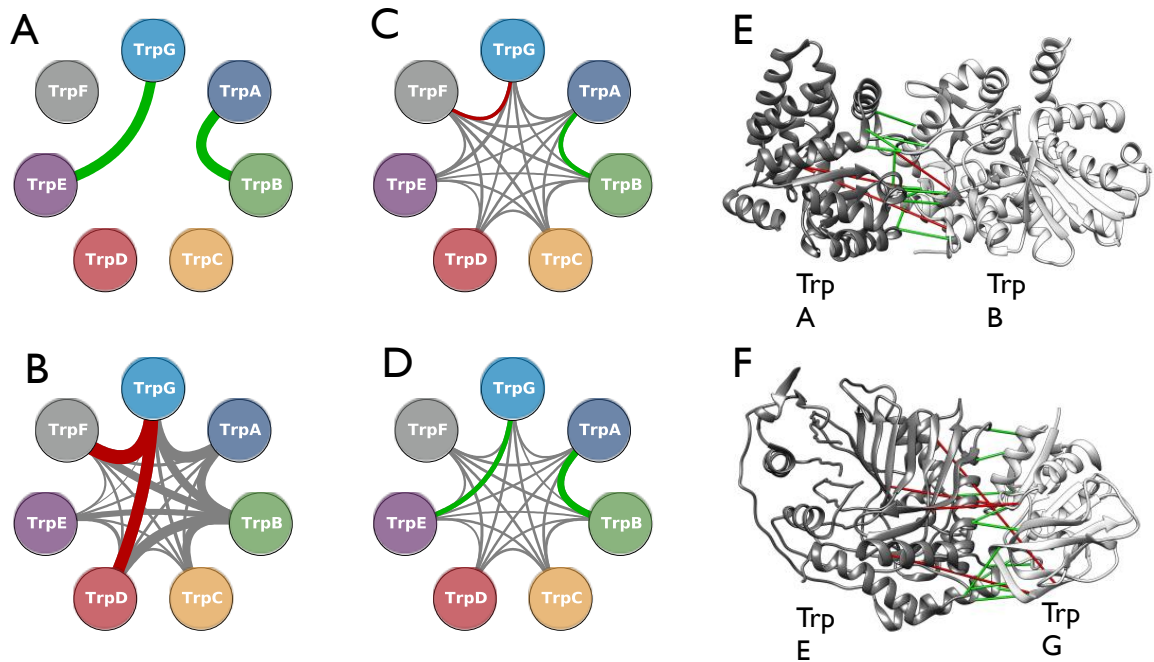

**Figure 4: Detection of protein-protein interactions between enzymes of the Tryptophan biosynthesis pathway.** **Panel A** shows the known PPI between the seven enzymes in the Trp pathway, only TrpA-TrpB and TrpE-TrpG are known to interact. **Panel B-D** show the results for the seed matching (matched by uniqueness in genome), for matchings of 1000 sequences per protein pair, for the full matching. Line width is proportional to the inter-protein coevolution score, the first two predictions are colored (TP = green, FP = red). For the seed alignment none of the true PPIs is recognized, for 1000 sequences one out of two. The second true PPI has the third score, but there is no gap between true and false PPI. For the full matching, the known PPI are found as the two highest-scoring, with scores detached from an almost continuous distribution of the remaining 19 scores. **Panel E-F** show the PDB structures of the complexes TrpA-TrpB and TrpE-TrpF, together with the 15 highest DCA-scoring inter-protein pairs, colored again in green for TP inter-protein contact predictions (12 for TrpA-TrpB, 11 for TrpE-TrpG) and in red for FP predictions (3 for TrpA-TrpB, 4 for TrpE-TrpG). The contact prediction is based on the fully matched PPM alignments.



\title{SUPPLEMENTARY INFORMATION  \\
\line(1,0){300}\\
Simultaneous identification of specifically interacting paralogs and inter-protein contacts by Direct-Coupling Analysis}

\author[1]{Thomas Gueudre}
\author[1,2]{Carlo Baldassi}
\author[1]{Marco Zamparo}
\author[3]{Martin Weigt}
\author[1,2]{Andrea Pagnani}

\affil[1]{Department of Applied Science and Technology, Politecnico di Torino, 10129 Torino, Italy}
\affil[2]{Human Genetic Foundation, Molecular Biotechnology Center, 10126 Torino}
\affil[3]{Sorbonne Universites, UPMC, Institut de Biologie Paris-Seine, CNRS, Laboratoire de Biologie Computationnelle et Quantitative UMR 7238, 75006 Paris France}

\date{}
\maketitle

\section{Gaussian Direct Coupling Analysis}

The basic steps for inferring contacts used in this work is the Direct
Coupling Analysis (DCA) \cite{weigt2009PNAS,marks2011direct,morcos2011direct}.
In this section, we recall the steps of inferring contact points,
given an already matched multiple sequence alignment (MSA) $A$ of
$M$ sequences of length $L$. Because the present
work requires the statistical model to be re computed several times,
we opted for the computationally fastest and simplest modeling, the
Multivariate Gaussian Modeling (MGM) introduced in \cite{Baldassi2014PLOSONE}

\subsection{Notation}

An MSA of $M$ sequences of length $L$ is represented by a $M\times L$~-~dimensional
array $A=\left(a_{i}^{m}\right)_{\,i\,=\,1,\,\ldots,\,L}^{\,m\,=\,1,\,\ldots,\,M}$,
where $a$ belongs to an alphabet of $Q+1=21$ symbols corresponding
to the $Q=20$ standard amino acids plus the ``gap'' symbol (-).
We transform the MSA into a $M\times\left(Q\cdot L\right)$ - dimensional
array $X=\left(x_{i}^{m}\right)_{i=1,\ldots,QL}^{m=1,\ldots,M}$ over
a binary alphabet $\left\{ 0,1\right\} $ accounting for the occupation
state of each residue position. Precisely, for $a=1,\dots,Q$ and
$l=1,\dots,L$, $x_{\left(l-1\right)Q+a}^{m}:=1$ if the standard
amino acid $a$ is present at residue position $l$, $x_{\left(l-1\right)Q+a}^{m}:=0$
otherwise. Notice that $x_{\left(l-1\right)Q+a}^{m}=0$ for all $a$
if the position $l$ corresponds to a gap ant that at most one of
the $Q$ variables $x_{\left(l-1\right)Q+1}^{m},\dots,x_{\left(l-1\right)Q+Q}^{m}$
can be equal to one. We denote the row length of $X$ by $N:=QL$.

The empirical covariance matrix $C=\left(C_{nn^{\prime}}\right)_{n,n^{\prime}=1,\dots,N}$
associated do the data $X$ is defined by:
\begin{equation}
C_{nn^{\prime}}:=\frac{1}{M}\sum_{m=1}^{M}\left(x_{n}^{m}-\bar{x}_{n}\right)\left(x_{n^{\prime}}^{m}-\bar{x}_{n^{\prime}}\right)\label{eq:covariance}
\end{equation}
where $\bar{x}_{n}:=\frac{1}{M}\sum_{m=1}^{M}x_{n}^{m}$ is the empirical
mean. We collect empirical means into the vector $\bar{x}=\left(\bar{x}_{n}\right)_{n=1,\dots,N}$.

Within the multivariate Gaussian model with normal-inverse-Wishart
prior introduced in Ref.~\cite{Baldassi2014PLOSONE}, the \emph{maximum
a posteriori} (MAP) estimation of the model mean $\mu$ is simply equal to:
\begin{equation}
\mu:=(1-\lambda)\bar{x}+\lambda \eta
\label{mu_def}
\end{equation}
and for covariance matrix $\Sigma$, given by:
\begin{equation}
\Sigma:=\lambda U+\left(1-\lambda\right)C+\lambda\left(1-\lambda\right)\left(\bar{x}-\eta\right)^{T}\left(\bar{x}-\eta\right)
\label{eq:covariance_with_pc}
\end{equation}

Here, $\lambda\in\left[0,1\right]$ is a parameter determining the
relative strength of the prior, which we call ``pseudocount'' and
which we typically set to the value $0.5$. The vector $\eta$ and
the matrix $U$ are the prior estimators of the mean and the covariances,
respectively. We take $\eta$ to be a uniform row vector of length
$N$ with entries all equal to $\left(Q+1\right)^{-1}$, and $U$
to be a $N\times N$~-~block-diagonal matrix composed of $Q\times Q$
blocks: the diagonal blocks have entries $Q\left(Q+1\right)^{-2}$
on the diagonal and $-\left(Q+1\right)^{-2}$ off-diagonal, while
the off-diagonal blocks are set to $0$.

With these definitions, the logarithm of the MAP value (log-MAP in
the following) for a multivariate Gaussian model is, up to an additive
constant:
\begin{equation}
\mathcal{L}=-\frac{1}{2}\log\det\Sigma\label{eq:logMAP}
\end{equation}

We refer to Ref.~\cite{Baldassi2014PLOSONE} for details.

\section{The Matching Problem}

\subsection{Matching definition \label{def_matching}}

Assume two alignment arrays $X^{1}$ and $X^{2}$ are given, whose
rows represent amino-acid sequences for two protein families in binary
encoding. The number of columns of the two matrices is $N_{1}$ and
$N_{2}$, respectively. We group the sequences (rows of the arrays)
in $S$ contiguous chunks, such that all sequences within a group
belong to the same species. At first, we assume that the number of
sequences for any given species is the same for the two families.
We denote by $M_{s}$ the number of sequences of the species $s$,
so that $\sum_{s=1}^{S}M_{s}=M$. Setting $B_{1}:=0$ and $B_{s}:=\sum_{k=1}^{s-1}M_{k}$
if $S>1$, all rows corresponding to the species $s$ have indices
in the interval $\mathcal{I}_{s}:=\left\{ B_{s}+1,\dots,B_{s}+M_{s}\right\} .$

Our objective is to find the correct association between the sequences
of the two families in each species by maximization of coevolution
signals. Given our assumptions, this association is a matching between
row $m$ of the array $X^{1}$ and a row $\pi\left(m\right)$ of the
array $X^{2}$, $\pi$ being a permutation of the row indices of $X^{2}$
which preserves the species: $\pi\left(m\right)\in\mathcal{I}_{s}$
if $m\in\mathcal{I}_{s}$. We denote by $X^{2\pi}$ the array $X^{2}$
with rows permuted according to $\pi$, i.e.~the elements of $X^{2\pi}$
are $\left(X^{2\pi}\right)_{mn}=X_{\pi\left(m\right)n}^{2}$. We also
denote by $X^{\pi}$ the $M\times N$~-~array, with $N:=N_{1}+N_{2}$,
obtained by concatenating $X^{1}$ and $X^{2\pi}$. Finally, we define
$C^{\pi}$, $\Sigma^{\pi}$ and $\mathcal{L}^{\pi}$ to be the empirical
covariance matrix, MAP covariance matrix and log-MAP, respectively,
associated to the concatenated data $X^{\pi}$ via the multivariate
Gaussian model relations, eqs.~\eqref{eq:covariance}, \eqref{eq:covariance_with_pc}
and \eqref{eq:logMAP}.

The two covariance matrices have a block structure:

\begin{eqnarray}
C^{\pi} & = & \left[\begin{array}{c|c}
C_{1} & D^{\pi}\\
\hline \left(D^{\pi}\right)^{T} & C_{2}
\end{array}\right]\\
\Sigma^{\pi} & = & \left[\begin{array}{c|c}
\Sigma_{1} & \Phi^{\pi}\\
\hline \left(\Phi^{\pi}\right)^{T} & \Sigma_{2}
\end{array}\right]\\
(\Sigma^{-1})^{\pi} & = & \left[\begin{array}{c|c}
\Sigma^{-1}_{1} & \Psi^{\pi}\\
\hline \left(\Psi^{\pi}\right)^{T} & \Sigma^{-1}_{2}
\end{array}\right]
\label{interact_blocks}
\end{eqnarray}

The diagonal parts (of sizes $N_{1}\times N_{1}$ and $N_{2}\times N_{2}$)
describe correlations within each protein, while the extra-diagonal
blocks $D^{\pi}$ and $\Phi^{\pi}$ (of size $N_{1}\times N_{2}$),
describe correlations (i.e.~possible co-evolution) between the two
proteins. The extra-diagonal blocks will then be the focus of our
proposed matching strategies.

\subsection{Scoring the matching}

Our strategy for finding the best matching consists in maximizing
some score within the multivariate Gaussian model for the joined families.
From a Bayesian perspective, $\pi$ is an additional latent variable
to infer, and one would ideally want to maximize the log-MAP of eq.~\eqref{eq:logMAP},
i.e.~to find an optimal matching $\pi^{\star}$ defined as:
\begin{equation}
\pi^{\star}:=\arg\max_{\pi}\left(\mathcal{L}^{\pi}\right)
\end{equation}

This is an arduous computational task: for realistic cases, the space
search is huge (it grows faster than exponentially), $\mathcal{L}^{\pi}$
is rather costly to compute, and the landscape while varying $\pi$
is especially rugged, such that classic search strategies such as
Simulated Annealing are infeasible. We thus resort to heuristic strategies,
which have good performance in practice.

A first observation is that we can consider alternatives to the score
function $\mathcal{L}^{\pi}$: according to the wisdom of co-evolution,
pairs of interacting proteins should exhibit some co-evolution signals,
encoded in both the covariance matrix $\Sigma^{\pi}$ and its inverse,
the interaction matrix $J^{\pi}=-\left(\Sigma^{\pi}\right)^{-1}$.
As their coefficients at $i$ and $j$ quantify the strength of coevolution
between site $i$ and $j$, one could maximize some quantity involving
$\Sigma^{\pi}$ or $J^{\pi}$. As the simplest choice, in the following
we will consider the squared Frobenius norm of the off-diagonal MAP
covariance, $\left\Vert \Phi^{\pi}\right\Vert _{F}^{2}=Tr\left(\left(\Phi^{\pi}\right)^{T}\Phi^{\pi}\right)$,
as an additional scoring function, besides $\mathcal{L}^{\pi}$.

A second observation is that, as we shall show below, for both these
scores we can devise a reasonably efficient hill climbing procedure:
starting from some initial matching $\pi_{0}$, we produce a sequence
of successive matchings $\pi_{1},\pi_{2},\dots$ by repeated application
of a local optimization scheme, such that at each step the score that
we are trying to maximize is (at least approximately) non-decreasing.

Finally, following the previous observation, we devised a strategy
for obtaining a better matching by ``mixing'' two or more sub-optimal
matchings.

In what follows, we detail the hill climbing procedure for the two
score functions and the mixing procedure. These will form the building
blocks of our overall heuristic strategies, which are explained in
the following section.

\subsubsection{Frobenius norm hill climbing}

We first study the Frobenius norm score, defined as:
\begin{eqnarray}
\left\Vert \Phi^{\pi}\right\Vert _{F}^{2} & = & Tr\left(\left(\Phi^{\pi}\right)^{T}\Phi^{\pi}\right)\\
 & = & \sum_{n=1}^{N_{1}}\sum_{n^{\prime}=1}^{N_{2}}\left(\Phi_{nn^{\prime}}^{\pi}\right)^{2}\nonumber 
\end{eqnarray}

The basic idea is to derive the basic step of a hill climbing procedure
as a local optimization process, in which one starts with a permutation
$\pi$ and tries to find a similar permutation $\pi^{\prime}$ which
maximizes the difference $\left\Vert \Phi^{\pi^{\prime}}\right\Vert _{F}^{2}-\left\Vert \Phi^{\pi}\right\Vert _{F}^{2}$.
To simplify the notation, let us rewrite the expression for the extra-diagonal
block of $\Sigma^{\pi}$ in eq.~\eqref{eq:covariance_with_pc} as:
\begin{equation}
\Phi^{\pi}=\left(1-\lambda\right)D^{\pi}+L
\end{equation}

so that we can write:
\begin{equation}
\left(\Phi_{nn^{\prime}}^{\pi^{\prime}}\right)^{2}-\left(\Phi_{nn^{\prime}}^{\pi}\right)^{2}=\left(1-\lambda\right)^{2}\left(\left(D_{nn^{\prime}}^{\pi^{\prime}}\right)^{2}-\left(D_{nn^{\prime}}^{\pi}\right)^{2}\right)+2\left(1-\lambda\right)\left(D_{nn^{\prime}}^{\pi^{\prime}}-D_{nn^{\prime}}^{\pi}\right)L_{nn^{\prime}}
\end{equation}

We now focus on the first addendum; first, we define the matrices
$Y^{1}$ and $Y^{2}$ obtained from $X^{1}$ and $X^{2}$ by subtracting
their mean, such as their elements are, for $n\in\left\{ 1,\ldots,N^{1}\right\} $,
$n^{\prime}\in\left\{ 1,\ldots,N^{2}\right\} $, $m\in\left\{ 1,\dots,M\right\} $:
\begin{eqnarray}
Y_{mn}^{1} & = & X_{mn}^{1}-\bar{x}_{n}\\
Y_{mn^{\prime}}^{2} & = & X_{mn^{\prime}}^{2}-\bar{x}_{N^{1}+n^{\prime}}\nonumber 
\end{eqnarray}
With these, we can write eq.~\eqref{eq:covariance} as: 
\begin{equation}
D_{nn^{\prime}}^{\pi}=\frac{1}{M}\sum_{m=1}^{M}Y_{mn}^{1}Y_{\pi\left(m\right)n^{\prime}}^{2}
\end{equation}
and therefore
\begin{eqnarray}
\left(O_{nn^{\prime}}^{\pi^{\prime}}\right)^{2}-\left(O_{nn^{\prime}}^{\pi}\right)^{2} & = & \frac{1}{M^{2}}\left(\left(\sum_{m=1}^{M}Y_{mn}^{1}Y_{\pi^{\prime}\left(m\right)n^{\prime}}^{2}\right)^{2}-\left(\sum_{m=1}^{M}Y_{mn}^{1}Y_{\pi\left(m\right)n^{\prime}}^{2}\right)^{2}\right)
\end{eqnarray}

We then restrict ourselves to permutations $\pi^{\prime}$ which only
differ within a single species $s$ from $\pi$:
\[
\pi\left(m\right)=\pi^{\prime}\left(m\right)\;\textrm{if}\:m\notin\mathcal{I}_{s}
\]
and obtain:
\begin{eqnarray}
\left(D_{nn^{\prime}}^{\pi^{\prime}}\right)^{2}-\left(D_{nn^{\prime}}^{\pi}\right)^{2} & = & \frac{2}{M^{2}}\sum_{m\in\mathcal{I}_{s}}\sum_{m^{\prime}\notin\mathcal{I}_{s}}Y_{mn}^{1}Y_{m^{\prime}n}^{1}\left(Y_{\pi^{\prime}\left(m\right)n^{\prime}}^{2}-Y_{\pi\left(m\right)n^{\prime}}^{2}\right)Y_{\pi\left(m^{\prime}\right)n^{\prime}}^{2}+\nonumber \\
 &  & +\mathcal{O}\left(\left(\frac{M_{s}}{M}\right)^{2}\right)
\end{eqnarray}
\begin{equation}
D_{nn^{\prime}}^{\pi^{\prime}}-D_{nn^{\prime}}^{\pi}=\frac{1}{M}\sum_{m\in\mathcal{I}_{s}}Y_{mn}^{1}\left(Y_{\pi^{\prime}\left(m\right)n^{\prime}}^{2}-Y_{\pi\left(m\right)n^{\prime}}^{2}\right)
\end{equation}
If we neglect the terms of order $\mathcal{O}\left(\left(\frac{M_{s}}{M}\right)^{2}\right)$,
i.e.~we assume that each single species has a few proteins compared
with the size of the dataset, we can write:
\begin{eqnarray}
\left\Vert \Phi^{\pi^{\prime}}\right\Vert _{F}^{2}-\left\Vert \Phi^{\pi}\right\Vert _{F}^{2} & \propto & \sum_{m\in\mathcal{I}_{s}}\sum_{n=1}^{N_{1}}\sum_{n^{\prime}=1}^{N_{2}}Y_{mn}^{1}\left(Y_{\pi^{\prime}\left(m\right)n^{\prime}}^{2}-Y_{\pi\left(m\right)n^{\prime}}^{2}\right)\left(\frac{\left(1-\lambda\right)}{M}\sum_{m^{\prime}\notin\mathcal{I}_{s}}Y_{m^{\prime}n}^{1}Y_{\pi\left(m^{\prime}\right)n^{\prime}}^{2}+L_{nn^{\prime}}\right)\nonumber \\
 & = & \sum_{m\in\mathcal{I}_{s}}\sum_{n=1}^{N_{1}}\sum_{n^{\prime}=1}^{N_{2}}Y_{mn}^{1}\left(Y_{\pi^{\prime}\left(m\right)n^{\prime}}^{2}-Y_{\pi\left(m\right)n^{\prime}}^{2}\right)\left(\Phi_{nn^{\prime}}^{\pi}-\frac{1}{M}\sum_{m^{\prime}\in\mathcal{I}_{s}}Y_{m^{\prime}n}^{1}Y_{\pi\left(m^{\prime}\right)n^{\prime}}^{2}\right)\nonumber \\
 & = & -\sum_{m\in\mathcal{I}_{s}}\mathcal{M}_{m\pi^{\prime}\left(m\right)}^{\pi,s}+\mathcal{A}^{\pi}\label{eq:FrobNorm_matching}
\end{eqnarray}

where in the last step we defined $\mathcal{A}^{\pi}$, which does
not depend on $\mathscr{\pi^{\prime}}$and is therefore irrelevant
for our optimization problem, and the cost matrix
\begin{eqnarray}
\mathcal{M}_{mm^{\prime}}^{\pi,k} & = & -\sum_{n=1}^{N_{1}}\sum_{n^{\prime}=1}^{N_{2}}Y_{mn}^{1}\left(\Phi_{nn^{\prime}}^{\pi}-\frac{1}{M}\sum_{m^{\prime\prime}\in\mathcal{I}_{s}}Y_{m^{\prime\prime}n}^{1}Y_{\pi\left(m^{\prime\prime}\right)n^{\prime}}^{2}\right)Y_{m^{\prime}n^{\prime}}^{2}\label{eq:FrobNorm_matching_costs}
\end{eqnarray}
Equation \eqref{eq:FrobNorm_matching} has the form of a matching
problem: we have an $M_{s}\times M_{s}$ cost matrix $\mathcal{M}^{\pi,s}$
and we want to find an optimal matching between the rows and the columns
indices, such that the sum of the costs is minimal (and therefore
the step in the Frobenius norm from $\pi$ to $\pi^{\prime}$ is maximal).
This problem is computationally easy and can be solved very efficiently
(e.g.~via linear programming), in particular for small $M_{s}$.

Therefore, if we start from any permutation $\pi$, we can derive
a new permutation $\pi^{\prime}$ by choosing a species $s$ and solving
a small matching problem; the new permutation will only differ on
the $s$-th block, and will likely have a bigger Frobenius norm, and
can serve as basis for further iterations. We call this algorithm
``Frobenius norm hill climbing''. The computation can be done reasonably
efficiently as it only requires linear algebra operations and solving
small matching problems; furthermore, in practice we only compute
the matrices $\mathcal{M}^{\pi,s}$ once for the whole dataset at
each iteration, after which we use them to update all the blocks independently
in parallel, and use the new permutation to compute new matrices $\mathcal{M}^{\pi^{\prime},s}$
and so on. This parallel method of update is not only useful to save
some computational time, but proves better in practice as a way to
avoid fixed points in the iterative algorithm. A pseudocode for this
procedure is shown in Algorithm~\ref{alg:FrobHillClimbing}.

The reason for using this algorithm is that it proved heuristically
to be very fast and efficient in the early stages of the optimization,
i.e.~when starting from a random permutation, as will be discussed
below.

\begin{algorithm}
\caption{\texttt{\label{alg:FrobHillClimbing}FrobNormHillClimbing}. This routine
implements the Frobenius norm hill climbing algorithm derived in the
text. Its arguments are the two alignment matrices minus the mean
($Y^{1}$ and $Y^{2}$ in the text, \texttt{Y1} and \texttt{Y2} here),
the list of blocks indices ($\left\{ \mathcal{I}_{s}\right\} _{s\in\left\{ 1,\ldots,S\right\} }$
in the text, \texttt{Ilist} here), an initial permutation ($\pi$
in the text, \texttt{permutation} here) and a group of parameters
(\texttt{pseudocount} and \texttt{iterations}). It returns a new permutation
($\pi^{\prime}$ in the text, \texttt{new\_permutation} here). Most
auxiliary routines (e.g.~\texttt{num\_columns}, \texttt{permute\_rows})
should have obvious meanings; \texttt{compute\_MAP\_covariance} implements
eq.~\eqref{eq:covariance_with_pc}; \texttt{permutation\_by\_matching}
calls some solver for the matching problem to obtain a permutation;
multiplication (denoted by \texttt{{*}}) is intended to be matrix
multiplication when matrices or vectors are involved; square brackets
are used to denote elements of lists and submatrices (e.g.~ \texttt{Y1{[}I,
\{1,...,N1\}{]}}, where \texttt{I} is a range, denotes a submatrix
obtained by taking the rows \texttt{I} of the matrix \texttt{Y1},
and all of its columns).}

\texttt{function FrobNormHillClimbing(Y1, Y2, Ilist, permutation,
pseudocount, iterations)}

\texttt{\{}

\texttt{~~N1 = num\_columns(Y1)}

\texttt{~~N2 = num\_columns(Y2)}

\texttt{~~M = num\_rows(Y1)}

\texttt{~~S = num\_elements(Ilist)}~\\

\texttt{~~range1 = \{1,...,N1\}}

\texttt{~~range2 = \{N1+1,...,N\}}

\texttt{~~new\_permutation = \{1,...,M\}}

\texttt{~~for iter = 1,...,iterations}

\texttt{~~\{}

\texttt{~~~~Y2p = permute\_rows(Y2, permutation)}

\texttt{~~~~Y = horizontal\_concatenation(Y1, Y2p)}

\texttt{~~~~C = compute\_MAP\_covariance(Y, pseudocount)}

\texttt{~~~~Phi = C{[}range1, range2{]}}

\texttt{~~~~for s = 1,...,S}

\texttt{~~~~\{}

\texttt{~~~~~~I = Ilist{[}s{]}}

\texttt{~~~~~~bY1 = Y1{[}I, \{1,...,N1\}{]}}

\texttt{~~~~~~bY2p = Y2p{[}I, \{1,...,N2\}{]}}

\texttt{~~~~~~T = Phi - ((1-pseudocount) {*} transpose(bY1)
{*} bY2p) / M}

\texttt{~~~~~~COSTS = bY1 {*} T {*} transpose(bY2)}

\texttt{~~~~~~new\_permutation{[}I{]} = permutation\_by\_matching(COSTS,
I)}

\texttt{~~~~\}}

\texttt{~~~~permutation = new\_permutation}

\texttt{~~\}}

\texttt{~~return new\_permutation}

\texttt{\}}
\end{algorithm}

\subsubsection{Log-MAP hill climbing \label{log_scores}}

Here, we perform a similar analysis to the one in the previous section
for the log-MAP score $\mathcal{L}^{\pi}$: we consider two permutations
$\pi$ and $\pi^{\prime}$, and we wish to some $\pi^{\prime}$ which
maximizes the difference
\begin{eqnarray}
\mathcal{L}^{\pi^{\prime}}-\mathcal{L}^{\pi} & =- & \frac{1}{2}\log\det\Sigma^{\pi^{\prime}}+\frac{1}{2}\log\det\Sigma^{\pi}
\end{eqnarray}
The concavity of the logarithm of the determinant of positive definite
matrices ensures that:

\begin{eqnarray}
\mathcal{L}^{\pi^{\prime}}-\mathcal{L}^{\pi} & \ge- & \frac{1}{2}\mathrm{Tr}\,\left(\Sigma^{\pi}\right)^{-1}\left(\Sigma^{\pi^{\prime}}-\Sigma^{\pi}\right)
\end{eqnarray}
therefore, we focus on minimizing only the term:
\begin{eqnarray}
\mathrm{Tr}\,\left(\Sigma^{\pi}\right)^{-1}\left(\Sigma^{\pi^{\prime}}-\Sigma^{\pi}\right) & = & \lambda\left(\left(\Sigma^{\pi}\right)^{-1}\right)\left(C^{\pi^{\prime}}-C^{\pi}\right)
\end{eqnarray}
Again, this can be written as a matching problem:
\begin{eqnarray}
\mathrm{Tr}\,\left(\Sigma^{\pi}\right)^{-1}\left(\Sigma^{\pi^{\prime}}-\Sigma^{\pi}\right) & \propto & \sum_{m=1}^{M}\mathcal{W}_{m\pi^{\prime}\left(m\right)}^{\pi}+\mathcal{B}^{\pi}\label{eq:logMAP_matching}
\end{eqnarray}
where $\mathcal{B}^{\pi}$ does not depend on $\pi^{\prime}$ and
is therefore irrelevant, and the matching weights are encoded in the
matrix: 
\begin{equation}
\mathcal{W}^{\pi}=Y^{1}\Psi^{\pi}\left(Y^{2}\right)^{T}\label{eq:logMAP_matching_weights}
\end{equation}
where we recall (Eq.\ref{interact_blocks}) that $\Psi$ is the extra diagonal block of $\Sigma^{-1}$.

We are only interested in the diagonal blocks of the matrix
$\mathcal{W}^{\pi}$, for which $m,m^{\prime}\in\mathcal{I}_{s}$
for some $s$, and we can perform the maximization independently and
in parallel for each species block, as for the Frobenius norm case.
In this way, we can define an iterative process which takes a given
permutation $\pi$ as input and produces a new permutation $\pi^{\prime}$
such that $\mathcal{L}^{\pi^{\prime}}\ge\mathcal{L}^{\pi}$, and therefore
produce a sequence of permutations with non-decreasing log-MAP. The
pseudocode for this procedure, which is even more computationally
efficient than the Frobeinus gradient ascent, is shown in Algorithm~\ref{alg:LogMAPHillClimbing}.

Unfortunately, our tests show that this algorithm, which we call ``Log-MAP
hill climbing'', is extremely prone to get trapped into fixed points.
For this reason, we mostly use this method for refinement of solutions
obtained by other means, since it typically does not provide big gains
in terms of log-MAP (both in terms of gain-per-iteration and in terms
of total gain up to the fixed point), even when starting from random
initial permutations.

\begin{algorithm}
\caption{\texttt{\label{alg:LogMAPHillClimbing}LogMAPHillClimbing}. This routine
implements the Log-MAP hill climbing algorithm derived in the text.
It's very similar to \texttt{FrobNormHillClimbing} (see Algorithm~\ref{alg:FrobHillClimbing}),
but runs until a fixed point is reached.}

\texttt{function LogMAPHillClimbing(Y1, Y2, Ilist, permutation, psudocount)}

\texttt{\{}

\texttt{~~N1 = num\_columns(Y1)}

\texttt{~~N2 = num\_columns(Y2)}

\texttt{~~M = num\_rows(Y1)}

\texttt{~~S = num\_elements(Ilist)}~\\

\texttt{~~range1 = \{1,...,N1\}}

\texttt{~~range2 = \{N1+1,...,N\}}

\texttt{~~new\_permutation = \{1,...,M\}}

\texttt{~~fixed\_point = false}

\texttt{~~while fixed\_point == false}

\texttt{~~\{}

\texttt{~~~~Y2p = permute\_rows(Y2, permutation)}

\texttt{~~~~Yp = horizontal\_concatenation(Y1, Y2p)}

\texttt{~~~~C = compute\_MAP\_covariance(Yp, pseudocount)}

\texttt{~~~~invC = inverse(C)}

\texttt{~~~~invPhi = invC{[}range1, range2{]}}

\texttt{~~~~for s = 1,...,S}

\texttt{~~~~\{}

\texttt{~~~~~~I = Ilist{[}s{]}}

\texttt{~~~~~~bY1 = Y1{[}I, \{1,...,N1\}{]}}

\texttt{~~~~~~bY2 = Y2{[}I, \{1,...,N2\}{]}}

\texttt{~~~~~~COSTS = bY1 {*} invPhi {*} transpose(bY2)}

\texttt{~~~~~~block\_permutation = permutation\_by\_matching(COSTS,
I)}

\texttt{~~~~~~new\_permutation{[}I{]} = block\_permutation}

\texttt{~~~~\}}

\texttt{~~~~if new\_permutation == permutation}

\texttt{~~~~\{}

\texttt{~~~~~~fixed\_point = true}

\texttt{~~~~\}}

\texttt{~~~~permutation = new\_permutation}

\texttt{~~\}}

\texttt{~~return new\_permutation}

\texttt{\}}
\end{algorithm}

\subsubsection{Mixing local optima}

As we mentioned above, the log-MAP hill climbing strategy shows a
strong tendency to get stuck in local maxima. Supposing that we have
obtained two different matchings $\pi_{1}$ and $\pi_{2}$ in such
way, e.g.~by initializing the algorithm from different initial random
configurations, a simple and effective way to improve over these solutions
is to obtain a new permutation $\pi^{\prime}$ by solving again the
matching problem defined by eq.~\eqref{eq:logMAP_matching}, in which
however the matching weight matrix is obtained by
\[
\mathcal{W}=\frac{1}{2}\left(\mathcal{W}^{\pi_{1}}+\mathcal{\mathcal{W}}^{\pi_{2}}\right)
\]
where the weights $\mathcal{W}^{\pi_{1}}$ and $\mathcal{W}^{\pi_{2}}$
are computed according to eq.~\eqref{eq:logMAP_matching_weights}.
The new permutation can then be refined via Log-MAP hill climbing.
The pseudocode for this procedure is shown in Algorithm~\ref{alg:MixPermutations}

This algorithm is generalizable in a number of ways (e.g.~we could
mix more than two solutions, tune the relative weights according to
the associated Log-MAP, etc.), but our empirical tests show that using
two permutations at a time seems to be the most effective approach.

\begin{algorithm}
\caption{\texttt{\label{alg:MixPermutations}MixPermutations}. This routine
implements the algorithm for mixing two permutations presented in
the text. It's very similar to \texttt{LogMAPHillClimbing} (see Algorithm~\ref{alg:LogMAPHillClimbing}),
but takes two input permutations ($\pi_{1}$ and $\pi_{2}$ in the
text, \texttt{p1} and \texttt{p2} here), and returns only one.}

\texttt{function MixPermutations(Y1, Y2, Ilist, p1, p2, pseudocount)}

\texttt{\{}

\texttt{~~N1 = num\_columns(Y1)}

\texttt{~~N2 = num\_columns(Y2)}

\texttt{~~M = num\_rows(Y1)}

\texttt{~~S = num\_elements(Ilist)}~\\

\texttt{~~range1 = \{1,...,N1\}}

\texttt{~~range2 = \{N1+1,...,N\}}

\texttt{~~Y2p1 = permute\_rows(Y2, p1)}

\texttt{~~Y2p2 = permute\_rows(Y2, p2)}

\texttt{~~Yp1 = horizontal\_concatenation(Y1, Y2p1)}

\texttt{~~Yp2 = horizontal\_concatenation(Y1, Y2p2)}

\texttt{~~C1 = compute\_MAP\_covariance(Yp1, pseudocount)}

\texttt{~~C2 = compute\_MAP\_covariance(Yp2, pseudocount)}

\texttt{~~invC1 = inverse(C1)}

\texttt{~~invC2 = inverse(C2)}

\texttt{~~invPhi1 = invC1{[}range1, range2{]}}

\texttt{~~invPhi2 = invC2{[}range1, range2{]}}

\texttt{~~invPhiMix = (invPhi1 + invPhi2) / 2}

\texttt{~~new\_permutation = \{1,...,M\}}

\texttt{~~for k = 1,...,S}

\texttt{~~\{}

\texttt{~~~~I = Ilist{[}k{]}}

\texttt{~~~~bY1 = Y1{[}I, \{1,...,N1\}{]}}

\texttt{~~~~bY2 = Y2{[}I, \{1,...,N2\}{]}}

\texttt{~~~~COSTS = bY1 {*} invPhiMix {*} transpose(bY2)}

\texttt{~~~~block\_permutation = permutation\_by\_matching(COSTS,
I)}

\texttt{~~~~new\_permutation{[}I{]} = block\_permutation}

\texttt{~~\}}

\texttt{~~return new\_permutation}

\texttt{\}}
\end{algorithm}

\section{Computational strategies}

We introduced the basic strategies for maximizing the two scoring
functions and mixing different sub-optimal solutions. We now outline
two different computational strategies for maximizing globally the
permutation $\pi$. We start first by outlining the \emph{Iterative
Paralog Matching}, our most accurate strategy with larger computational complexity.
Then, we outline the progressive paralog matching strategy, which turns out
to be marginally less accurate then the Iterative
Paralog Matching, but with
a much lower computational complexity.

\subsection{Iterative
Paralog Matching \label{sec:Consensus-Method}}

We describe here the complete Iterative
Paralog Matching which we
used to derive the results presented in the main text. It uses all
three computational building blocks of the previous section; as an
additional, final heuristic pass, it also employs a refinement aimed
once again at escaping local maxima.

The protocol starting point is the generation of a large number of
random permutations. Each of those is then used as a starting point
for a Frobenius norm hill climbing phase. These are all independent
and thus can be run in parallel. The number of iterations during this
phase is a parameter of the protocol; we observed that in practice
a plateau is typically reached after about $10$ iterations. In the
following phase, we perform log-MAP hill climbing up to a fixed point
(which is normally reached in a short number of iterations), again
in parallel and independently for each configuration. After this,
we collect all these configurations in a set, and we rank them according
to their log-MAP score. We then apply this procedure iteratively:
we take the two lowest-ranking configurations, removing them from
the set; we mix them as described above, and obtain a new (typically
better then both) configuration; we add this new configuration to
the set. This phase continues until there is only one configuration
left. In the final phase, we try to optimize futher this result by
the following procedure: given a configuration, we produce a number
(e.g.~32) of partially scrambled versions of it, and then we mix
them progressively as in the previous phase, until we end up again
with a single configuration. The scrambling is performed in this way:
we fix a fraction (e.g.~50\%) of all the matching indices, and randomize
the rest while keeping the condition that interactions are only allowed
within each species. This procedure is intended to escape from local
maxima, and is iterated until it is judged that it is no longer effective
(in our tests, this happened after $100$ to $200$ iterations).

Of course, this protocol can be improved in many ways. In fact, we
also developed a simpler, trivially parallelizable and incremental
version, in which the final phase is avoided and the mixing is performed
by taking random pairs of configurations (thus avoiding the ranking).
This protocol had similar performances in terms of the maximum value
of the log-MAP that it was able to reach, at the cost of requiring
a much larger number of initial configurations to start with.

A simplified pseudo-code for our protocol is shown in \ref{alg:FindAlignment}.

\begin{algorithm}
\caption{\label{alg:FindAlignment}\texttt{FindAlignment}. This algorithm
implements the complete optimization protocol described in the text.
Its arguments are the two alignment matrices minus the mean ($Y^{1}$
and $Y^{2}$ in the text, \texttt{Y1} and \texttt{Y2} here), the list
of blocks indices ($\left\{ \mathcal{I}_{s}\right\} _{s\in\left\{ 1,\ldots,S\right\} }$
in the text, \texttt{Ilist} here), a number of permutations to start
with (\texttt{num\_initial\_permutations}), a number of permutations
to use in the last phase (\texttt{num\_scrambled\_permutations}),
and some parameters (\texttt{pseudocount}, \texttt{frob\_iterations},
\texttt{final\_phase\_iterations}, \texttt{scrambling fraction}).
It calls \texttt{FrobGradientAscent} (see Algorithm~\ref{alg:FrobHillClimbing})
to bootstrap from random iterations and then iteratively mixes pairs
of permutations with \texttt{MixPermutations} (see Algorithm~\ref{alg:LogMAPHillClimbing})
according to their ranking (auxiliary function \texttt{mix\_permlist\_ranked},
defined here) until only one permutation remains. The final pass scrambles
the permutation, producing a new list which is then reduced again
via mixing. \texttt{LogMAPGradientAscent} (see Algorithm~\ref{alg:MixPermutations})
is used as a refinement after each step.}

\texttt{function FindAlignment(Y1, Y2, Ilist, num\_initial\_permutations,
num\_scrambled\_permutations,}

\texttt{~~~~~~~~~~~~~~~~~~~~~~~pseudocount,
frob\_iterations, final\_phase\_iterations,}

\texttt{~~~~~~~~~~~~~~~~~~~~~~~scrambling\_fraction)}

\texttt{\{}

\texttt{~~perm\_list = generate\_random\_permutations(num\_initial\_permutations,
Ilist)}

\texttt{~~for i = 1,...,num\_elements(perm\_list)}

\texttt{~~\{}

\texttt{~~~~new\_perm = FrobGradientAscent(Y1, Y2, Ilist, permlist{[}i{]},
pseudocount, iterations)}

\texttt{~~~~new\_perm = LogMAPGradientAscent(Y1, Y2, Ilist, new\_perm,
pseudocount)}

\texttt{~~~~perm\_list{[}i{]} = new\_perm}

\texttt{~~\}}

\texttt{~~final\_perm = mix\_permlist\_ranked(Y1, Y2, Ilist, permlist,
pseudocount)}

\texttt{~~for t = 1,...,final\_phase\_iterations}

\texttt{~~\{}

\texttt{~~~~perm\_list = generate\_scrambled\_perms(final\_perm,
num\_scrambled\_perms, scrambling\_fraction)}

\texttt{~~~~final\_perm = mix\_permlist\_ranked(Y1, Y2, Ilist,
permlist, pseudocount)}

\texttt{~~\}}

\texttt{~~return final\_perm}

\texttt{\}}

~

\texttt{function mix\_permlist\_ranked(Y1, Y2, Ilist, permlist, pseudocount)}

\{

\texttt{~~while num\_elements(perm\_list) > 1}

\texttt{~~\{}

\texttt{~~~~perm\_list = sort\_by\_logMAP(perm\_list, Y1, Y2,
pseudocount)}

\texttt{~~~~p1 = perm\_list{[}end{]}}

\texttt{~~~~p2 = perm\_list{[}end-1{]}}

\texttt{~~~~new\_perm = MixPermutations(Y1, Y2, Ilist, p1, p2,
pseudocount)}

\texttt{~~~~new\_perm = LogMAPGradientAscent(Y1, Y2, Ilist, new\_perm,
pseudocount)}

\texttt{~~~~perm\_list{[}end-1{]} = new\_perm}

\texttt{~~~~perm\_list = drop\_last\_element(perm\_list)}

\texttt{~~\}}

\texttt{~~return perm\_list{[}1{]}}

\texttt{\}}
\end{algorithm}

\subsection{Progressive Paralog Matching}

The Iterative
Paralog Matching outlined in Section~\ref{sec:Consensus-Method},
as discussed in the main text, turns out to be extremely accurate
in terms of reproducing the correct matching on the two-component
system biological dataset. However, due to computational complexity
issues, it hardly scales for genome-wide analysis. To overcome such
limitation, we propose a faster and simpler heuristic strategy: the
Progressive Paralog Matching. 

Due to the size of the matching space, we propose
a step-by-step inference strategy by including larger and larger chunks
(i.e.~block of species) to the alignments to be matched. To proceed
recursively, we need to single out, at each step, the matching with
the greatest likelihood, employing an Maximum A Posteriori Estimator
(MAP). The criterion to select a given species $s$ is the entropy
$\omega_{s}$, defined as the log of the number of possible matchings
of homologs \textit{within} this genome. Considering now the general case in
which the species sizes can be different for different families, we
denote by $M_{s}^{1}$ (resp. $M_{s}^{2}$) the number of protein
sequences in species $s$ found in the alignment of protein family
$\mathcal{F}_{1}$ (resp. $\mathcal{F}_{2}$). Assuming, for example,
that $M_{s}^{1}>M_{s}^{2}$:

\begin{equation}
\omega_{s}=\log\left(\frac{M_{s}^{1}!}{\left(M_{s}^{1}-M_{s}^{2}+1\right)!}\right)
\label{entropic_def}
\end{equation}

We are going to denote $X^{\omega}$ and $C^{\omega}$ the data and
correlation matrices obtained by matching all species $s$ characterized
by an entropy $\omega_{s}\le\omega$. 

\textit{Initialization step}: Genomes readily matched by uniqueness
($M_{s}^{1}=M_{s}^{2}=1$), have an entropy $\omega_{s}=0$ and therefore
provide a natural initialization $X^{0},C^{0},\Sigma^{0}$. 

\textit{Propagation step}: We then proceed recursively. We assume that the matching is \textit{known} for species up to entropy less than or equal to $\omega$. The model inferred given that matching has parameters $(\mu,\Sigma)$. We consider the next species, say $q$, of entropy immediately above $\omega$, $\omega_{q}>\omega$. The set of sequences in $q$ defines two sub-MSA for family $1$ and $2$, $X^{1}$ and $X^{2}$. As explained in Sec.\ref{def_matching}, a matching $\pi$ is defined as a concatenation of $X^2$ on $X^1$. We denote by $X^{\pi}$ the full, concatenated, MSA.

$X^{\pi}$ having a small number of rows w.r.t the whole dataset, it only slightly perturbs the empirical correlation matrix $C^{\pi}=C+\Delta C$, and similarly $\Sigma^{\pi}=\Sigma+\Delta \Sigma$ from Eq.\ref{eq:covariance_with_pc}.
At this point, the same reasoning presented in Sec.\ref{log_scores} can be used. More precisely, one can score the best sub-matching $\pi$ for species $q$ by evaluating the score matrix:
\begin{equation}
\mathcal{W}=\left(X^{1}-\mu^1 \right) \Psi^{-1}\left(X^{2}-\mu^2\right)^{T}\\
\end{equation}
with $\mu^1$ ($\mu^2$) respectively the $N_1$ first (the $N_2$ last) components of the mean vector $\mu$, and $\Psi$ the extra-diagonal block of $\Sigma^{-1}$ (Eq.\ref{interact_blocks}).
Once the cost matrix is computed, the best matching $\pi$ can be recovered by standard linear programming. The newly matched species $q$ is added to the pool of known species, and the new model parameters $(\mu,\Sigma)$ recomputed by adding the block $X^{\pi}$ to $X^{\omega}$. 

The above step is repeated until the full alignment is matched. This algorithm is very scalable, as it runs over an alignment of $20000$ sequences in less than $10$ minutes, on a laptop (implementation in Julia).

\subsection{Contact Map Predictions and PPI DCA Scoring}

All contact predictions presented in the Main Text, such as Fig.2, 3 and 4 are done by Pseudo-Likelihood Maximization \cite{Ekeberg2013}, using the Julia Package (\url{github.com/pagnani/PlmDCA}) with default parameters. 

The scoring of the interactions between the Tryptophan proteins as presented in Fig.4 was done using the procedure described in \cite{Feinauer2016}: we ranked the inter-protein scores from the largest, and consider the mean over the $4$ largest.  We also checked other scores found in the literature \cite{ovchinnikov2014robust}; they do not change the conclusion of the study.

\section{Statistics Tables about the Tryptophan Dataset}

\begin{figure*}[htb!]
   \includegraphics[width=0.9\textwidth]{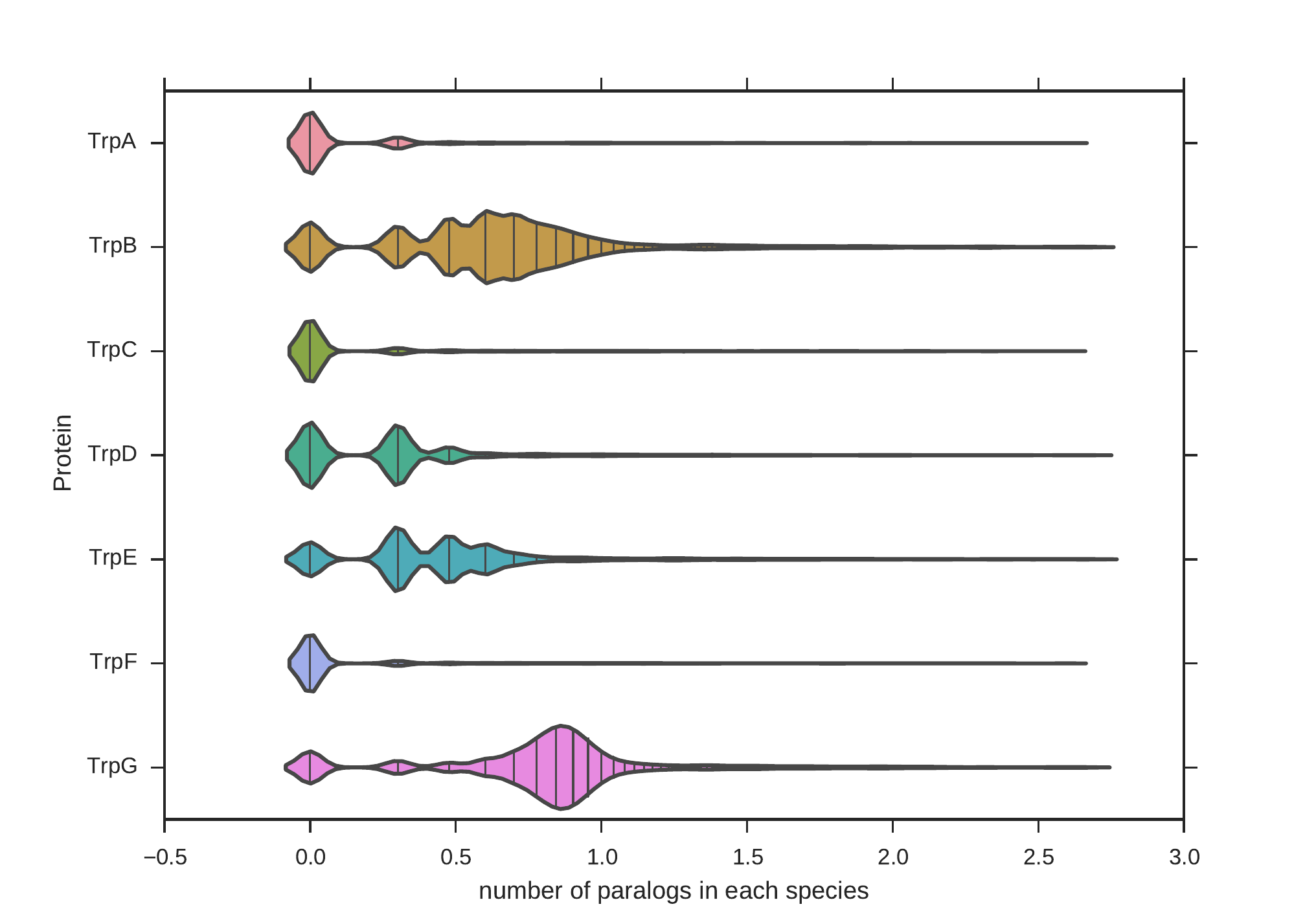} 
    \caption{The distribution of the log10 number of paralogs in species, for each Trp protein. A cut-off at a maximum of $500$ paralogs by species has been chosen for plotting convenience. The sticks represent the boxes of the histogram used to generate the smooth violin plot. The elongated structures come from few species that present a very high number (greater than $100$) of paralogs.}
    \label{violin_plot}
\end{figure*}

\begin{figure*}[htb!]
   \includegraphics[width=1.1\textwidth , trim=5cm 0 7cm 0cm,angle=90]{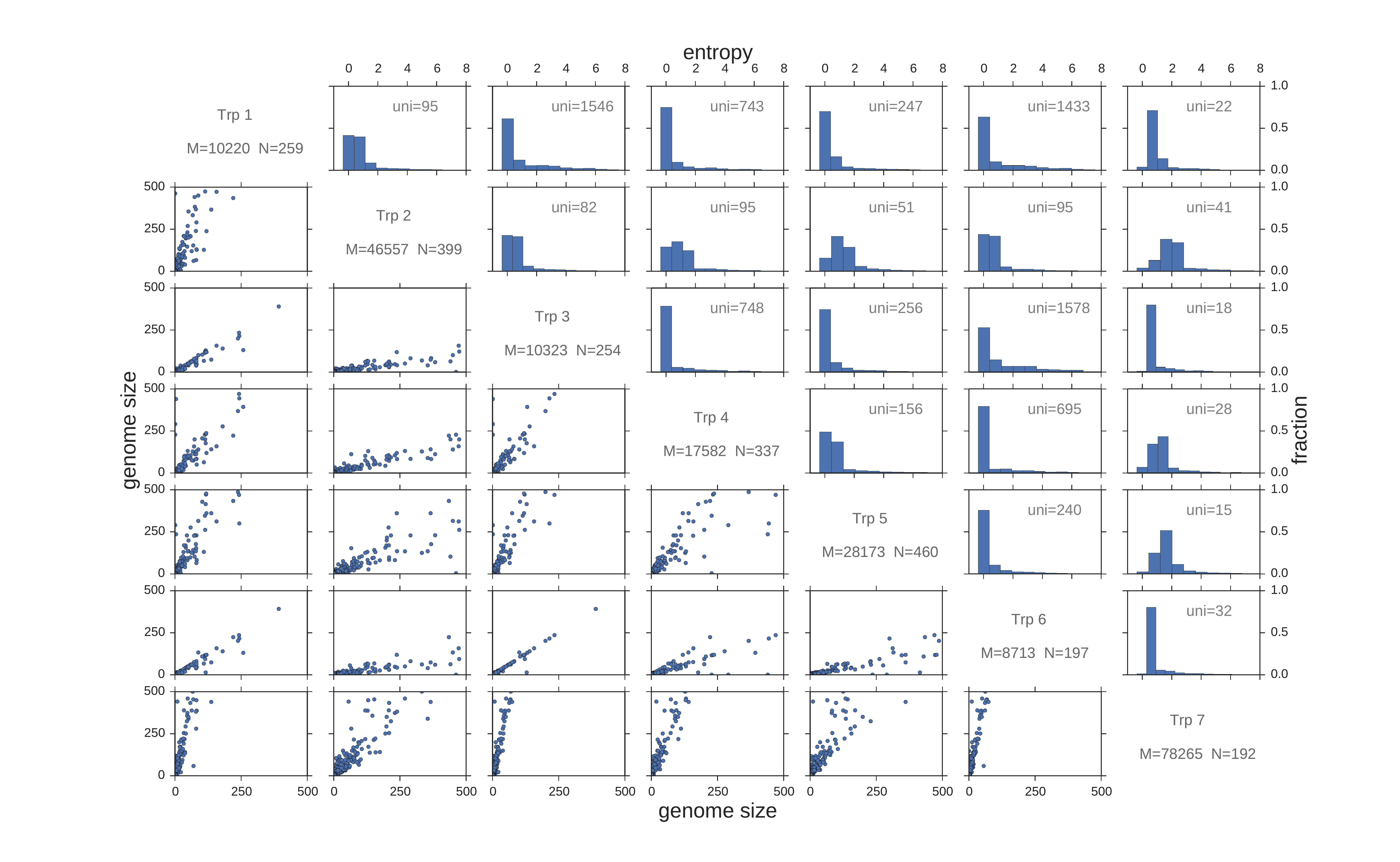} 
    \caption{Table of histograms. The above triangular part represents the histogram of the entropies of the species, as given Eq.\ref{entropic_def}. $uni$ is the size of the matched by uniqueness alignments. The diagonal part recalls the statistics presented in Table.\ref{trp_singleSizes}. Finally the lower triangular part are scatter plots: each point represents a particular species, with its coordinates as the number of paralogs for each protein. A cut-off of $500$ paralogs by species has been chosen for plotting convenience.}
    \label{EntropicplotTrip}
\end{figure*}

The following table contains various statistics about the set of Tryptophan alignments used to assess the interaction network. The set is made of seven proteins, labelled from A to G. Table \ref{trp_singleSizes} contains information about the single proteins. Instead, Table \ref{trp_pairSizes} contains the statistics of resulting matched pairs of alignments using various methods: uniqueness, genetic or from co-evolution. Finally, Figs.\ref{violin_plot} and \ref{EntropicplotTrip} present a more complete overview of the paralogs statistics of this dataset.

\begin{table}
\begin{center}
\begin{tabular}{|c|c|c|c|c|c|}
\hline
 & L & M & P & S & Quartiles \\
\hline
TrpA & 259 & 10220 & 4.457 & 32.604 & (1.0,1.0,2.0) \\ 
TrpB & 399 & 46557 & 16.992 & 145.826 & (3.0,4.0,6.0) \\ 
TrpC & 254 & 10323 & 4.536 & 39.868 &(1.0,1.0,1.0)  \\ 
TrpD & 337 & 17582 & 7.130 & 59.693 & (1.0,2.0,2.0) \\ 
TrpE & 460 & 28173 & 11.749 & 124.933 & (2.0,3.0,4.0) \\ 
TrpF & 197 & 8713 & 4.122 & 32.400 & (1.0,1.0,1.0) \\ 
TrpG & 192 & 78265 & 24.713 & 187.331 & (5.0,7.0,9.0) \\ 
\hline
\end{tabular}
\end{center}
\caption{For each protein in the Tryptophan Operon, the size of the protein $L$, the total number of sequences $M$ in the alignments. $P$ indicates the average number of paralogs per species and $S$ the standard deviation. Finally, the three quartiles, in order, are presented in the last column. More details are given in Fig.\ref{violin_plot}.}
\label{trp_singleSizes}
\end{table}

\begin{table}
\begin{center}
\begin{tabular}{|c|c|c|c|c|c|}
\hline
 &  & unique & genetic & covariation & score \\
\hline
TrpA & TrpB & 95 & 4374 & 4915 & 0.337 \\ 
TrpA & TrpC & 1546 & 3198 & 6255 & 0.137 \\ 
TrpA & TrpD & 743 & 2823 & 6188 & 0.121 \\ 
TrpA & TrpE & 247 & 3118 & 5285 & 0.115 \\ 
TrpA & TrpF & 1433 & 3357 & 5701 & 0.139 \\ 
TrpA & TrpG & 22 & 4646 & 4176 & 0.118 \\ 
TrpB & TrpC & 82 & 3326 & 4425 & 0.145 \\ 
TrpB & TrpD & 95 & 3737 & 7242 & 0.112 \\ 
TrpB & TrpE & 51 & 3911 & 10720 & 0.096 \\ 
TrpB & TrpF & 95 & 3643 & 4064 & 0.137 \\ 
TrpB & TrpG & 41 & 8053 & 16437 & 0.090 \\ 
TrpC & TrpD & 748 & 3392 & 5778 & 0.129 \\ 
TrpC & TrpE & 256 & 2976 & 4839 & 0.127 \\ 
TrpC & TrpF & 1578 & 3825 & 5811 & 0.135 \\ 
TrpC & TrpG & 18 & 4272 & 3827 & 0.135 \\ 
TrpD & TrpE & 156 & 2681 & 7469 & 0.100 \\ 
TrpD & TrpF & 695 & 2819 & 5165 & 0.149 \\ 
TrpD & TrpG & 28 & 6249 & 6450 & 0.129 \\ 
TrpE & TrpF & 240 & 2519 & 4295 & 0.113 \\ 
TrpE & TrpG & 15 & 5324 & 9796 & 0.245 \\ 
TrpF & TrpG & 32 & 3635 & 3457 & 0.126 \\ 
\hline
\end{tabular}
\end{center}
\caption{For all possible pairings of proteins, the resulting size of the concatenated MSA, following various matching procedures: \textit{unique} for matched by uniqueness; \textit{genetic} for matched by genetic distance; \textit{covariation} for matched by co-evolution analysis. \textit{score} is the score obtained from the alignments matched by co-evolution analysis, as presented in \cite{Feinauer2016}.}
\label{trp_pairSizes}
\end{table}

 \bibliographystyle{unsrt}
\bibliography{biblio}

\end{document}